\newcommand{\be}{\begin{equation}}
\newcommand{\ee}{\end{equation}}
\newcommand{\bea}{\begin{eqnarray}}
\newcommand{\eea}{\end{eqnarray}}

\newcommand{\bra}[1]{\left\langle #1 \right|}
\newcommand{\ket}[1]{\left| #1 \right\rangle}
\def\slash#1{\setbox0=\hbox{$#1$}#1\hskip-\wd0\dimen0=5pt\advance
       \dimen0 by-\ht0\advance\dimen0 by\dp0\lower0.5\dimen0\hbox
         to\wd0{\hss\sl/\/\hss}}

\def\HH{{\cal H}}
\def\VV{{\cal V}}
\def\AA{{\cal A}}
\begin{document}
\setlength{\baselineskip}{3.0ex}
\vspace*{1cm}
\rightline{BARI-TH/93-156}
\vspace*{3.5cm}
\begin{center}
{\large\bf Theoretical Approaches to $B \to K^* \gamma$ \\}
\vspace*{6.0ex}
{\large\rm G. Nardulli \\}
\vspace*{1.5ex}
{\large\it Dipartimento di Fisica, Univ. di Bari\\
I.N.F.N., Sezione di Bari, Italy}
\end{center}
\vspace*{4cm}
\noindent
\begin{quotation}
\vspace*{3cm}
\begin{center}
\begin{bf}
Abstract
\end{bf}
\end{center}
\vspace*{5mm}
\noindent
We describe two recent theoretical calculations of the $B \to K^* \gamma$
decay. The first method is based on an effective lagrangian
incorporating chiral symmetry for the light degrees of freedom
and the heavy quark spin-flavor symmetries for the heavy quarks. Using
these symmetries, together with the hypothesis of polar
q$^2$ dependence of the semileptonic form factors, one is able to relate
$B \to K^* \gamma$ and $D \to K^* \ell \nu$ decays with the result
$ R \, =\,  \Gamma(B \to K^* \gamma) / \Gamma(b \to s \gamma) = 0.09 \pm 0.03$.
The second method is based on 3-point function QCD sum rules and gives the
result $ R \, = \, 0.17 \pm 0.05$. Both results are in reasonable agreement
with recent data
from the CLEO II Collaboration.
\end{quotation}
\newpage
\section{Introduction}

Radiative $B$-meson decays of the type $B \to K^*(892) \gamma$
have been extensively studied in the last few years \cite{MASIERO}
since they are known to provide, through loop effects, interesting information
on some of the Standard Model parameters: Kobayashi Maskawa
matrix elements and the top quark mass. More recently the CLEO II
Collaboration has published data on this decay \cite{CLEO}, thus
renewing the interest in the subject. The aim of this talk is to review
two recent calculations of the radiative rare B decay based respectively on
an effective lagrangian in the infinite heavy quark mass limit
\cite{noi0} and on 3-point QCD sum rules \cite{colangelo}.

The electromagnetic penguin operator describing the
$b \to s \gamma$ transition is given,
for $m_s << m_b$, by:
\be \HH_{eff} = C m_b \epsilon^{\mu *} {\bar s} \sigma_{\mu \nu}
{(1 + \gamma_5) \over 2} q^\nu b \hskip 3 pt , \label{eq1}\ee
\par\noindent
where $\epsilon$ and $q$ are the photon polarization and momentum.
The constant $C$, neglecting $m_c$, is given by:
\be C = {G_F \over \sqrt 2} {e \over 4 \pi^2} \; V^*_{ts} \; V_{tb} \;
F_2( { {m_t^2} \over {m_W^2}}) \label{eq2}\ee

\noindent where the function $F_2$ has
been computed in \cite{MASIERO}, \cite{INAMI}
and depends weakly on the top quark mass: in the range
$90\; GeV < m_t <  210 \; GeV$ it increases
from $0.55$ to $0.68$. This leads  to the prediction,
using $m_b=4.6 \;GeV$, $\tau_B=1.4 \; ps$ and $m_t=120 \; GeV$,
of the inclusive $b \to s \gamma$ branching ratio:
$BR(b \to s \gamma) = 2.2 \; (|V_{ts}|/0.042)^2 \times 10^{-4}$.

\par
Let us now consider the $B \to K^* \gamma$ decay.
The amplitude for
$B(p) \to K^*(p',\eta) \; \gamma(q,\epsilon)$
can be written as follows:
\be \AA(B \to K^* \gamma) = \bra{K^*(p',\eta)} J^\mu_{eff} \ket {B(p)} \hskip
3pt \epsilon_\mu^*
\label{eq5}\ee

\noindent
with
\be J^\mu_{eff} = C m_b \bar s  \sigma^{\mu \nu} {(1 + \gamma_5)\over 2}
q_\nu b \hskip 5pt. \label{eq6}
\ee

\noindent
 The
matrix element in
eq.(\ref{eq5}) contains one form factor $F_1$ defined
as follows ($q=p-p'$):
\bea
\bra{K^*(p',\eta)} \bar s \sigma_{\mu \nu} {(1 + \gamma_5) \over 2}
q^\nu b \ket {B(p)}
& = & [ i \epsilon_{\mu \nu \rho \sigma}
\eta^{*\nu} p^\rho p^{\prime \sigma}
\nonumber \\
&+& \frac{1}{2} [\eta^*_\mu (m^2_B - m^2_{K^*}) -
(\eta^* \cdot q) (p + p^\prime)_\mu]]  \, F_1(q^2)\; ,
 \label{eq7}\eea

\noindent
which has to be computed at the kinematical point $q^2=0$.

\section{Effective Lagrangian Approach}
Let us now describe the effective field approach
to $B \to K^* \gamma$. It is based on the implementation in
an effective lagrangian of the chiral symmetry for
the light quark  ($q= u, d, s$) degrees of freedom
and the spin-flavor symmetries for heavy quarks $Q$. We shall
work in the $m_Q= m_b$ or $m_c \to \infty$ limit, even though the role
of mass corrections may be relevant and certainly deserves further study.

We can introduce effective fields for
light and heavy mesons as follows.
The $J^P=0^-$
and $1^-$ heavy $Q{\bar q}_a$ mesons are represented by a $4\times 4 $
Dirac matrix \cite{wise}, \cite{noi1}:
\bea
H_a &=& \frac{(1+\slash v)}{2}[P_{a\mu}^*\gamma^\mu-P_a\gamma_5]\\
{\bar H}_a &=& \gamma_0 H_a^\dagger\gamma_0
\eea
where $a=1,2,3$ (for $u$, $d$, $s$), $v$ is the heavy meson velocity,
$P^{*\mu}_a$ and $P_a$ are annihilation operators satisfying
$\langle 0|P_a| H_a (0^-)\rangle  =\sqrt{m_H}$ and
$\langle 0|P^{*\mu}_a| H_a (1^-)\rangle  =  \epsilon^{\mu}\sqrt{m_H}$,
where $v^\mu P^*_{a\mu}=0$ and $m_H$ is the heavy meson mass.
Also the heavy positive parity mesons with
$J^P=0^+$
and $1^+$ can be included by a doublet $S$ \cite{noi1} defined
similarly to $H$ (with an extra $\gamma_5$).
The light pseudoscalar mesons are described by
\be
\xi=\exp{\frac{iM}{f_{\pi}}}
\ee
where $f_{\pi}=132 MeV$ and  ${M}$
is the usual $3 \times 3$ matrix describing the
octet of pseudoscalar Nambu-Goldstone bosons.
Under the chiral group $SU(3)\otimes SU(3)$
the fields transform as
follows
\bea
\xi & \to  & g_L\xi U^\dagger=U\xi g_R^\dagger \; ; \; \; H  \to   H
U^\dagger\\
\Sigma =\xi^2 & \to  & g_L\Sigma {g_R}^\dagger \; ; \; \; \;
{\bar H}  \to  U {\bar H}
\eea
where  $g_L$, $g_R$ are global $SU(3)$
transformations and $U$ depends on the space point $x$, the fields, $g_L$
and $g_R$.
\par
The vector meson resonances belonging to the low lying $SU(3)$ octet can
be introduced by using the hidden gauge symmetry approach,
where the $1^-$ particles are the
gauge bosons of a gauge local subgroup $SU(3)_{loc}$. Starting from chiral
$U(3)\otimes U(3)$, one gets a nonet of vector fields
$\rho_\mu$ that describe the particles $\rho$, $\omega$, $K^*$ and
$\phi$.
In this way the heavy quark effective chiral lagrangian describing the
fields $H$, $\xi$, $\rho_\mu$ as well as their interactions
can be easily constructed \cite{noi1}.
Since higher derivative terms would induce new coupling constants,
they are neglected, which however forces to consider kinematical
regimes where the light meson momenta are small.

Let us now discuss currents.
At the lowest order in the derivatives of the pseudoscalar fields, the weak
tensor current between light pseudoscalar and negative parity heavy mesons
is as follows
\be
L_{\mu\nu}^a=i\frac{\alpha}{2}\langle\sigma_{\mu\nu}(1+\gamma_5)H_b
\xi^\dagger_{ba}\rangle
\ee
that has the same transformation properties of the quark current
${\bar q}^a\sigma^{\mu\nu}(1+\gamma_5)Q$. Together with (11) we also consider
the weak effective current \cite{wise}, \cite{noi1} corresponding to the
quark $V-A$ current ${\bar q}^a\gamma^\mu(1-\gamma_5)Q$, i.e.
\be
L_\mu^a=i\frac{\alpha}{2}\langle\gamma_\mu(1-\gamma_5)H_b\xi^\dagger_{ba}
\rangle
\ee
One puts the same coefficient $i\alpha/2$ in (11) and (12)
because, as a
consequence of the equations of motion of the heavy quark,
one has, in the $b$ rest frame
\cite{isgur},
\be
\gamma^0 b=b \; .
\ee
Therefore
\be
{\bar q}^a\sigma_{0i}(1+\gamma_5)Q=-i{\bar q}^a\gamma_i(1-\gamma_5)Q
\ee
and the effective currents $L_{\mu\nu}^a$ and $L_\mu^a$ have to satisfy,
in the heavy meson rest frame, the relation
\be
L_{0i}=-iL_i
\ee
\par
One also introduces the weak tensor current containing the light vector meson
$\rho^\alpha$ and reproducing ${\bar q}^a\sigma^{\mu\nu}(1+\gamma_5)Q$
\be
L_{1a}^{\mu\nu}=i\alpha_1\left\{g^{\mu\alpha}g^{\nu\beta}-\frac{i}{2}
\epsilon^{\mu\nu\alpha\beta}\right\}\langle\gamma_5H_b\left[
\gamma_\alpha(\rho_\beta-\VV_\beta)_{bc}-\gamma_\beta(\rho_\alpha-
\VV_\alpha)_{bc}\right]\xi^\dagger_{ca}\rangle
\ee
where
$\VV_{\mu ba}=\frac{1}{2}\left(\xi^\dagger\partial_\mu \xi
+\xi\partial_\mu \xi^\dagger\right)_{ba}$.
$L_{1a}^{\mu\nu}$ is related to the vector current $L_{1a}^\mu$ introduced
in \cite{noi1} to represent ${\bar q}^a\gamma^\mu(1-\gamma_5)
Q$ between light vector particles and heavy mesons:
\be
L_{1a}^\mu=\alpha_1\langle\gamma_5 H_b(\rho^\mu-\VV^\mu)_{bc}\xi^\dagger_{ca}
\rangle
\ee
In a similar way currents containing the positive parity mesons
can be introduced.
\par
Using previous equations in connection with a
polar diagram with $1^+$ or $1^- \; B^*_s$
intermediate state between the current and the $B \, K^*$ system,
this approach gives rise to
a simple prediction for $F_1(q^2_{max})$, where $q^2_{max}=(m_B -m_{K^*})^2$.
In order to go to the physical point $q^2=0$ one makes the
further assumption of polar $q^2$ behaviour of the form factor. The final
result can be written in terms of the semileptonic form factors. Indeed one
obtains the relation:
\be
F_1(q^2) =  \left\{\frac{q^2+m_B^2-m_{K^*}^2}{2 m_B}
\frac{V(q^2)}{m_B+m_{K^*}}-\frac{m_B+m_{K^*}}{2 m_B}A_1(q^2)\right\}
\label{eq20} \ee
\\
 among the form
factor $F_1$ responsible for the transition $B \to K^* \gamma$
and the semileptonic form factors $V(q^2)$ and $A_1(q^2)$ defined by:
\bea
\langle K^*(p^\prime,\eta)|\bar s\gamma_{\mu}(1-\gamma_5)b|{\bar B}
(p)\rangle  &=& \frac{2 V(q^2)}{m_B+m_{K^*}}
\epsilon_{\mu\nu\lambda\sigma} \eta^{*\nu} p^\lambda p^{\prime\sigma}
\nonumber\\
\nonumber \\
&+& i (m_B + m_{K^*}) A_1(q^2) \eta^*_{\mu} +...
\label{eq21}\eea
\\
($\bar B= B^-$ or ${\bar B}^0$); in (\ref{eq21})
the ellipses denote terms proportial to $(p + p^{\prime})_\mu$
or $q_\mu$. The results are $V(0)=0.61 \pm 0.12, \, A_1(0)=0.20 \pm 0.02$.
In order to get $F_1(0)$ one has to note that the results
for $V(0)$ and $A_1(0)$ are obtained in this approach from the
analogous quantities of the decay $D \to K^* \ell \nu$ \cite{noi1} under
the hypotheis of polar $q^2$ behaviour of all the form factors. Since
$V$ and $A_1$ scale differently: $V(q^2_{max})/A_1(q^2_{max})\approx m_Q$,
it follows that $A_1$ in eq.(18) is a higher order correction and
should be neglected. The final result one gets is therefore
$F_1(0)=0.25 \pm 0.05$ and

\be R \, = \, {\Gamma(B \to K^* \gamma) \over \Gamma(b \to s \gamma)}
= \Big( {m_B \over m_b} \Big)^3 \; \Big(1 - {m^2_{K^*} \over m^2_B} \Big)^3 \;
|F_1(0)|^2 = 0.09 \pm 0.03 \hskip 5pt.\label{eq16b} \ee

\noindent
Using the computed value of the inclusive branching ratio,
one finds
$ BR(B \to K^* \gamma) =(2.0 \pm 1.2) \times 10^{-5}$ where
uncertainties in $R$ and $V_{ts}$ have been added
in quadrature. This result should be compared to the experimental
data from the CLEO II Collaboration \cite{CLEO}:
\be BR(B \to K^*(892) \; \gamma) = (4.5 \pm 1.5 \pm 0.9) \times 10^{-5} \;.
\label{eq0}\ee

\section{QCD Sum Rules Approach}

To compute $B \to K^* \gamma$ by QCD sum rules \cite{colangelo}, one
considers the three-point function correlator:
\be T_{\alpha \mu \nu}(p, p^\prime, q) = (i)^2 \int dx \; dy \;
e^{i(p^\prime \cdot x - p \cdot y)}
\bra {0} T (J_\alpha(x) \hat O_{\mu \nu} (0) J^\dagger_5 (y) ) \ket {0}
\label{eq8} \ee

\noindent
where the currents $J_\alpha$ and $J_5$
have the same quantum numbers of $K^*$ and $B$ mesons, respectively, i.e.
$J_\alpha (x)  =  \bar q(x) \gamma_\alpha s(x)$ and
 $J_5 (y)  =  \bar q(y) i \gamma_5 b(y) \hskip 5pt$, while the operator
$\hat O_{\mu \nu}(0)$ is given by
$\hat O_{\mu \nu} (0) = \bar s (0) {1\over 2} \sigma_{\mu \nu} b(0)$.
Using the decomposition
\be T_{\alpha \mu \nu} q^\nu = i \epsilon_{\alpha \mu \rho \sigma} p^\rho
p^{\prime \sigma}  T(p^2, p^{\prime 2}, q^2) \label{eq11} \ee

\noindent one can compute the invariant amplitude $T$ in QCD through the
Operator Product Expansion
for $p^2, p^{\prime 2}$ large and spacelike,
obtaining a perturbative
term and non-perturbative power corrections
 parameterized in terms of quark and gluon condensates.

The invariant amplitude $T (p^2, p^{\prime 2}, q^2)$ can be related,
by means of a
double dispersion relation, to a
 hadronic spectral density which gets contributions
from the lowest lying resonances, the $B$ and $K^*$ mesons, plus higher
resonances and a continuum:
\bea T^H (p^2, p^{\prime 2}, 0) & = & f_B { m_B^2 \over m_b} f_{K^*} m_{K^*}
{ 1 \over (p^2 - m^2_B) (p^{\prime 2} - m^2_{K^*}) } F_1(0)
\nonumber\\
\nonumber \\
& + &
higher\; resonances + continuum.
\label{eq13} \eea

\noindent In eq.(\ref{eq13}), $f_{K^*}$ and $f_{B}$
are the leptonic decay constants of the $K^*$ and
$B$ mesons, respectively, defined by:
$\bra{ 0} J_\mu \ket {K^*(p^\prime, \eta)} = f_{K^*} m_{K^*} \eta_\mu$,
and $\bra{ 0} J_5 \ket {B(p)} = f_B { m^2_B / m_b}$.
The value of $f_{K^*}$ can be obtained from the decay
$\tau^- \to K^{*-} \nu_\tau$:
$f_{K^*}=0.22\pm0.01 \; GeV$; as for $m_b$ and $f_B$ one uses
$m_b=4.6 \;GeV$  and, consequently,
$f_B=0.18 \pm 0.01 \;GeV$ as computed in
\cite{noi}.

According to duality, the higher resonance states and the continuum
 contribution
to the hadronic spectral function can be modelled by perturbative QCD.
A prediction for $F_1(0)$ can then be obtained
by equating the hadronic and the QCD sides of the sum rule
in a duality region where the quark behaviour is well
reproduced by a sum over infinitely many
resonances. The resulting Borel improved
sum rule can be found in \cite{colangelo}.
The numerical outcome is as follows:
$F_1(0)=0.35 \pm 0.05$ (the error is
obtained by varying the phenomenological parameters in their allowed
intervals).
This result allows us to estimate the fraction of inclusive
$b \to s \gamma$ decays
represented by the exclusive channel $B \to K^* \gamma$:

\be R \, = \, {\Gamma(B \to K^* \gamma) \over \Gamma(b \to s \gamma)}
= 0.17 \pm 0.05 \hskip 5pt.\label{eq16a} \ee
\section{Conclusions and outlook}

The two methods
give results that are in reasonable agreement (taking into
account theoretical uncertainties) with each other and
with the experimental data from the CLEO II Collaboration. They however
disagree in one point.
In deriving the values of $A_1$ and $V$ from $D \to K^* \ell \nu$,
the experimental analyses are performed by
assuming for the $q^2$ dependence
of the form factors a simple pole behaviour; it follows that in eq.(18)
$A_1(0)$ is subleading and the dominant contribution arises from the
$B_s^* \; 1^-$ pole (i.e. from the $V(0)$ term). On the other hand,
QCD sum rules seem to indicate, both in the $m_Q \to \infty$ limit \cite
{colangelo} and including all the mass corrections \cite{Ball} that
$A_1$ is almost constant in $q^2$, which implies that
both $A_1$ and $V$ should contribute to $F_1(0)$. It would be interesting
to reanalyze semileptonic data on the decay  $D \to K^* \ell \nu$
by assuming for the form factors the $q^2$ dependence given by
QCD sum rules. This would allow to obtain new values for
$A_1(0)$ and $V(0)$ for the $D \to K^*$ transition and, using
heavy flavour symmetry, for the $B \to K^*$ transition as well. In this way
a more careful comparison between the two
approaches would be feasible.
\par\noindent
\vspace*{1cm}
\par\noindent
{\bf Acknowledgements}
\par\noindent
It is a pleasure to thank R. Casalbuoni, C. A. Dominguez, P. Colangelo,
A. Deandrea, N. Di Bartolomeo, R. Gatto and N. Paver for most
fruitful discussions.

\end{document}